\documentclass{aa}  
\usepackage{graphicx}
\usepackage{txfonts}
\usepackage{colortbl}

\begin{document} 

\title{Mode coupling coefficients between
  the convective core and radiative envelope of $\gamma\,$Doradus
and  slowly pulsating B stars}


\author{C.\ Aerts\inst{1,2,3}
\and S. Mathis\inst{4}}

\institute{Institute of Astronomy, KU\,Leuven, Celestijnenlaan 200D,
  B-3001
  Leuven, Belgium\\
email:\ {Conny.Aerts@kuleuven.be}
\and
Department of Astrophysics, IMAPP, Radboud University Nijmegen, PO Box 9010,
6500 GL, Nijmegen, The Netherlands
\and 
Max Planck Institute for Astronomy, Koenigstuhl 17, 69117, Heidelberg, Germany
\and
Universit\'e Paris-Saclay, Universit\'e Paris Cit\'e, CEA, CNRS, AIM,
91191 Gif-sur-Yvette, France}

\date{Received June 15, 2023; Accepted July 25, 2023}

\abstract
    {Signatures of { coupling between an inertial mode in the
        convective core and a gravito-inertial mode in the envelope}
      have been found in four-year {\it Kepler\/} light curves of 16
      rapidly rotating $\gamma\,$Doradus ($\gamma\,$Dor) stars. This makes it possible to obtain a
      measurement of the rotation frequency in their convective
      core. { Despite their similar internal structure and
        available data, inertial modes have not yet been reported
        for slowly pulsating B (SPB) stars.}}
      {We aim to provide a numerical counterpart of the recently
        published theoretical { expressions for the mode-coupling
          coefficients, $\varepsilon$ and $\tilde{\varepsilon}$. These
          coefficients represent the two cases of a continuous and a
          discontinuous Brunt-V\"ais\"al\"a frequency profile at the
          core-envelope interface, respectively.  We consider
          $\gamma\,$Dor and SPB stars to shed
          light on the difference between these two classes of
          intermediate-mass gravito-inertial mode pulsators in terms
          of core and envelope mode coupling.}}
      {We used asteroseismic forward  models of two samples consisting
        of 26 SPB stars and 37 $\gamma\,$Dor stars
        to infer their numerical values of $\varepsilon$ { and
          $\tilde{\varepsilon}$.  For both samples, we also} computed: the linear
        correlation coefficients between $\varepsilon$ { or
          $\tilde{\varepsilon}$} and the near-core rotation frequency,
        the chemical gradient, the evolutionary stage, the convective
        core masses and radii, and the Sch\"onberg-Chandrasekhar
        limiting mass representing the maximum mass of an inert helium
        core at central hydrogen exhaustion that can still withstand
        the pressure of the overlaying envelope.}
      {The asteroseismically inferred values of $\varepsilon$
{ and $\tilde{\varepsilon}$ for the two samples}
are between 0.0 and 0.34. While
        $\varepsilon$ is most strongly correlated with the near-core
        rotation frequency for $\gamma\,$Dor stars, the fractional
        radius of the convective core instead provides the tightest
        correlation for SPB stars. We find
        $\varepsilon$ to decrease mildly as the stars evolve.
        { For the SPB stars,
          $\varepsilon$ and $\tilde{\varepsilon}$ have similar moderate
          correlations with respect to the core properties. For
          the $\gamma\,$Dor stars, $\tilde{\varepsilon}$ reveals systematically lower and often no
          correlation to the core properties; their  $\varepsilon$  is
          mainly determined by the near-core rotation frequency.}          
        The Sch\"onberg-Chandrasekar limit is already surpassed by the
        more massive SPB stars, while none of the
        $\gamma\,$Dor stars have reached it yet.  }
      {Our asteroseismic results for the mode coupling
        support the theoretical
        interpretation and reveal that young, fast-rotating
        $\gamma\,$Dor stars are most suitable for undergoing 
        couplings between inertial modes in the rotating convective
        core and gravito-inertial modes in the radiative
        envelope. The phenomenon has been found in 2.4\% of such
        pulsators with detected period spacing patterns, whereas it has not been seen in any
        of the SPB stars so far.}

\titlerunning{Mode coupling coefficients between core and envelope of
  $\gamma\,$Dor and SPB stars}
\authorrunning{Aerts et al.}

\maketitle
%

\section{Introduction}

Rotation is an important ingredient of stellar evolution models
\citep{Maeder2009}. However,  our
understanding of the physical processes inside stars induced by their
internal rotation is marked by some lingering deficiencies.  Thanks to asteroseismology, we have access to a
tool for measuring the internal profile from so-called rotational splitting of a
star's non-radial mode frequencies \citep{Ledoux1951}. This was first
turned into practice for the pressure modes of the Sun
\citep{Deubner1979} and subsequently for gravity modes in a white
dwarf \citep{Winget1991}, as well as for low-order pressure and gravity modes in a
young massive B-type dwarf \citep{Aerts2003,Dupret2004}.

{ The study of internal stellar rotation became an established
  field of research once the photometric light curves assembled with
  the NASA {\it Kepler\/} telescope reached a sufficiently long
  duration to resolve rotationally split frequencies. This led to
  estimates for the core rotation frequency from split dipole mixed
  modes in red giants
  \citep[e.g.][]{Beck2012,Mosser2012,Deheuvels2012,Beck2014,Gehan2018}
  and in subgiants
  \citep[e.g.][]{Deheuvels2014,Deheuvels2020}. Rotationally split
  multiplets also allowed to deduce the near-core rotation frequency
  of main sequence stars
  \citep[e.g.][]{Kurtz2014,Saio2015,Moravveji2016,SchmidAerts2016,
    VanReeth2016,GangLi2019,GangLi2020}.  A few more detections of
  internal rotation for multiple intermediate- and high-mass stars
  have been done with the BRITE constellation as well
  \citep[e.g.][]{Sowicka2017,Kallinger2017}.  The refurbished version
  of the {\it Kepler\/} project  (K2) subsequently delivered
  internal rotation measurements of white dwarfs \citep{Hermes2017},
  while those of subdwarfs are summarised in \citet{Charpinet2018}.
  Applications of rotation inversions delivered the entire rotation
  profile inside various types of slowly rotating stars, instead of
  just the (near-) core and envelope values
  \citep{Deheuvels2014,Deheuvels2015,Triana2015,DiMauro2016,Triana2017,Bazot2019}.
  \citet{Aerts2019} provided an overarching summary of all internal
  rotation measurements and discussed them in terms of stellar
  evolution and angular momentum transport mechanisms.  }

A particularly challenging case to measure internal stellar rotation
from the splitting of nonradial mode frequencies occurs when the modes
have frequency values comparable to the frequency of rotation. In such
a case, the Coriolis acceleration cannot be treated as a small
perturbative effect with respect to the other acting forces in the
computations of the eigenmodes of the star. A proper treatment
of the nonradial oscillation modes in such a situation can be done
adopting the traditional approximation of rotation
\citep[TAR,][]{LeeSaio1987,LeeSaio1997,Townsend2003,Mathis2009,Bouabid2013}.
The TAR is an excellent approximation for modes with spin parameters
$s\,\equiv\ 2\Omega/\omega\geq 1$ 
with $\Omega$ the rotation frequency and $\omega$ the frequency of the
mode. It is valid in the regime where $2\Omega<N$
and $\omega<\!\!<N$ with $N$ as the
{ Brunt-V\"ais\"al\"a (BV)} frequency
and in stars not flattened too much by
the centrifugal acceleration
\citep{Ouazzani2017,Dhouib2021a,Dhouib2021b,Henneco2021}, provided that the
horizontal displacement caused by the mode is dominant over the
vertical one. This latter condition is fulfilled for gravity modes of
stars in the core hydrogen burning phase of their evolution, as
proven by long-term high-resolution time-series spectroscopy of slowly
pulsating B (SPB) stars
\citep{Aerts1999,Mathias2001,DeCatAerts2002,Briquet2003,DeCat2005} and
$\gamma\,$Doradus ($\gamma\,$Dor) stars
\citep{Mathias2004,Aerts2004,DeCat2006}. Space asteroseismology
meanwhile showed that these conditions are also well met for most of
the gravito-inertial and Rossby modes active in the rapidly rotating
radiative envelope of intermediate- and high-mass pulsating dwarfs,
revealing spin parameters roughly in the range $s\in[10,30]$
\citep{Neiner2012b,Neiner2012a,Aerts2017,Saio2018}.

Methods for measuring the internal rotation profile, $\Omega(r)$, of
gravito-inertial mode pulsators have been developed since the four-year
light curves assembled with the NASA {\it Kepler\/} space telescope
allowed for the identification of series of such modes with
consecutive radial order
\citep{VanReeth2015,VanReeth2016,Ouazzani2017,Christophe2018,VanReeth2018,GangLi2019,GangLi2020,Takata2020}. Applications
of these methods have led to rigorous and consistent estimates of the
internal rotation frequency for hundreds of intermediate-mass stars
\citep[see ][for a summary]{Aerts2021}. Most of the
asteroseismic measurements deliver the internal rotation frequency at
the bottom of the radiative envelope, where a boundary layer occurs
between the convective core and the radiative envelope. The
gravito-inertial mode kernels have their strongest probing power
in that boundary layer
\citep{VanReeth2016,Ouazzani2017,Pedersen2018,Michielsen2019,Michielsen2021,Mombarg2021,Vanlaer2023}.
For a fraction of the pulsators with gravito-inertial modes, an
envelope or surface estimate of the rotation is also available
from pressure modes and rotational modulation, respectively
\citep{VanReeth2018,GangLi2020,Sekaran2021}.

Gravity and super-inertial gravito-inertial modes with
$\omega>2\Omega$  do not propagate in the convective
core of dwarfs \citep[e.g.][]{Prat2016,Prat2018} 
and can hence not deliver their core rotation
\citep[e.g.][for early attempts from rotation inversion for a
  $\gamma\,$Dor and SPB star, respectively]{Kurtz2014,Triana2015}.  A
major breakthrough on this front was achieved by \citet{Ouazzani2020},
who came up with a way to determine the rotation in the convective
core from inertial modes restored purely by the Coriolis
acceleration. Their theoretical study showed convincingly that
coupling between a pure inertial mode trapped in the convective core
and a sub-inertial gravito-inertial mode (with $\omega<2\Omega$)
propagating in the surrounding radiative
envelope may occur for models of rapidly rotating $\gamma\,$Dor stars.
\citet{Ouazzani2020} found such mode coupling to result in a
clear dip signature at particular frequencies in period spacing
patterns of envelope gravito-inertial modes.  Their theoretical study
was inspired by the predictions of \citet{Saio2018} to explain
observed period spacing patterns in the young rapidly rotating
$\gamma\,$Dor pulsator KIC\,5608334. To interpret the asteroseismic
data of this star, \citet{Saio2018} compared predictions based on the
TAR with computations taking full account of the Coriolis acceleration
from the method developed by \citet{LeeBaraffe1995} and found the
latter to reveal a dip structure in the period spacing diagram, while
the TAR predictions do not.

The findings by \citet{Ouazzani2020} then led \citet{Saio2021} to
revisit the {\it Kepler\/} data of $\gamma\,$Dor stars and study those
with a single clean dip signal in their period spacing pattern from
the perspective of coupling between inertial core modes and
gravito-inertial envelope modes.  They found 16 $\gamma\,$Dor stars
with such mode coupling and deduced the rotation frequency in their
convective core. This gives slightly faster rotation in the core than
in the envelope at a level of 10\% differentiality or less. They
also found an anticorrelation between the level of differentiality and
the evolutionary stage of the pulsators expressed as the hydrogen mass
fraction of the core, $X_c$, which ranges from 0.7 to 0.2 for the
16 pulsators with mode coupling.

Following onto \citet{Ouazzani2020} and \citet{Saio2021}, a deeper
theoretical understanding of the dip structure in period spacing
patterns { of $\gamma\,$Dor stars}
and of { their} 10\% level of differentiality between the core and
envelope rotation was offered by \citet{Tokuno2022}. Their study also
provides theoretical expressions for a dimensionless parameter,
$\varepsilon$, which they assume to attain values between 0 and 1 and
expressing a level of optimal circumstances for inertial modes in the
core to couple to sub-inertial gravito-inertial modes in the envelope.

{ While a dip structure in mode period spacings due to coupling
  between inertial core modes and gravito-inertial envelope modes has been
  well established in several $\gamma\,$Dor stars, no such signature
  has been reported yet for SPB stars. However, the inner structure of
  $\gamma\,$Dor and SPB stars is similar in the sense that they have a
  well-developed convective core surrounded by an often rapidly rotating
  boundary layer that connects the core to a radiative envelope.
The members of both classes of gravito-inertial pulsators cover the
entire range from slow rotation to almost critical rotation.
It is not yet understood why the members of these two classes of}
pulsators reveal period spacing patterns
with somewhat different morphological properties. Indeed, while {
  many of} the
$\gamma\,$Dor stars have long period spacing patterns involving tens
of modes of consecutive radial order, often with  just one dip or none at all
\citep{VanReeth2015,Keen2015,Bedding2015,GangLi2019,GangLi2020,Sekaran2021},
{ most of} the SPB stars have shorter patterns with oscillatory behaviour and/or
multiple dips
\citep{Degroote2010,Papics2012,Papics2014,Szewczuk2015,Papics2017,Szewczuk2018,Szewczuk2021,Pedersen2021}.
The detected periodicity in the dip pattern of
{ slow to moderate rotators among the $\gamma\,$Dor and
SPB stars is relatively well
understood in terms of a strong $\mu-$gradient ($\nabla_{\mu}$) due to
a receding convective core as the star evolves, causing strong mode
trapping \citep{Kurtz2014,Saio2015,SchmidAerts2016,Murphy2016,Pedersen2018,Michielsen2019,GangLi2019,Wu2020,Michielsen2021}.}
This phenomenon caused by $\nabla_{\mu}$, or by buoyancy glitches in
general, gives a specific morphology to period spacing patterns that
also occurs in the absence of fast rotation, as studied analytically
by \citet{Miglio2008} and \citet{Cunha2019}.
{ However, some} $\gamma\,$Dor and SPB stars reveal periodic and rather shallow
dips intertwined with one or a few sharp dips, which may point to the
simultaneous occurrence of { mode trapping due to
a strong $\mu-$gradient and} coupling between inertial and
gravito-inertial modes.

Moreover,
the presence of a core magnetic field may also give rise to a dip
structure, even when the TAR is used
\citep{Prat2019,Prat2020,VanBeeck2020,Dhouib2022}. Being able to
unravel the physical causes of all the observed dips thus offers the
future potential to detect and measure internal magnetic field
profiles in addition to the rotation profiles and check if they are
consistent with the predictions { for the internal magnetic field
strengths}
by \citet{Aerts2021-GIW}.
This requires a way to { unravel the signature of}
mode coupling from { the one due to $\nabla_{\mu}$, in the presence
  of rapid rotation and possible magnetic fields}
when modeling observed dips.
With this goal in mind,
{ we aim to provide a numerical value for the coupling coefficient}
by relying on the best asteroseismic models of a sample
of gravito-inertial pulsators consisting of both $\gamma\,$Dor and SPB
stars. { We wish to }
investigate
if numerical values of the parameters $\varepsilon$ and
$\tilde{\varepsilon}$
introduced by
\citet{Tokuno2022} from asteroseismic models provide a useful
prediction about the occurrence or absence of core-to-envelope mode
coupling. { In particular, our goal is to investigate whether
  $\gamma\,$Dor stars and SPB stars have different or similar values for
  $\varepsilon$ and $\tilde{\varepsilon}$. We will also
test if the values of $\varepsilon$ and
$\tilde{\varepsilon}$
}
indeed
occur in the interval $[0,1]$ as assumed by \citet{Tokuno2022} and
whether they are correlated with typical properties of the convective
core { of $\gamma\,$Dor and SPB stars, which have quite different
  size and mass. Finally, we will also search for relationships
  between the coupling coefficients and 
  the hydrogen mass fraction (as a good
proxy for the evolutionary stage) or the properties of $\mu$ and
$\nabla_{\mu}$ in the boundary layer between the convective core and
the radiative envelope for both types of pulsators.}

\section{Asteroseismic inference of the coupling coefficient, $\varepsilon$}

The theoretical work by \citet{Tokuno2022} defined a new parameter,
$\varepsilon$, capturing the level of opportunity for an 
interaction between a pure
inertial mode in the rotating convective core and a sub-inertial gravito-inertial
mode propagating in the rotating radiative envelope.  We therefore
call $\varepsilon$ the ``coupling coefficient'' (\citet{Tokuno2022}
did not provide any terminology for this parameter).  In their
analytical expressions derived for $\varepsilon$, \citet{Tokuno2022}
relied on some assumptions, one being weak interaction between the
oscillation modes in the core and envelope, such that $0 < \varepsilon
<\!\!<1$. They found $\varepsilon$ to decrease from 0.343 at stellar
birth to 0.018 near the terminal age main sequence of a $\gamma\,$Dor
star model of 1.5\,M$_\odot$ with solar metallicity and rotating at
2.2\,d$^{-1}$ (25.44$\mu$Hz) computed by \citet{Saio2021}.

The theoretical work by \citet{Tokuno2022} provided a physical
understanding of observed mode couplings in $\gamma\,$Dor stars. It
triggered our current study with the aim to derive values of
$\varepsilon$ for concrete gravito-inertial pulsators.  Following up
on \citet{Aerts2021-GIW}, we compute such numerical values of
$\varepsilon$ for the 63 gravito-inertial pulsators in that study,
relying on their best forward asteroseismic models. These were
computed by \citet{Mombarg2021} for the 37 $\gamma\,$Dor stars and by
\citet{Pedersen2021} for the 26 SPB stars using statistical
methodology inspired by \citet{Aerts2018}.  While some additional
g-mode pulsators have been modeled in the literature \citep[see][for
  a review]{Aerts2021}, the adopted methods are too diverse to add
them to the sample. Moreover, their numerical seismically calibrated
models are not available to us. Thus, we  restricted this study to
the homogeneously treated sample of 63 pulsators already considered in
\citet{Aerts2021-GIW}. Two $\gamma\,$Dor stars in this sample reveal
the envisioned coupling between an inertial core and a
gravito-inertial envelope mode.

\subsection{Internal profiles at the convective core boundary}

\begin{figure*}[t!]
\begin{center} 
\rotatebox{270}{\resizebox{6.5cm}{!}{\includegraphics{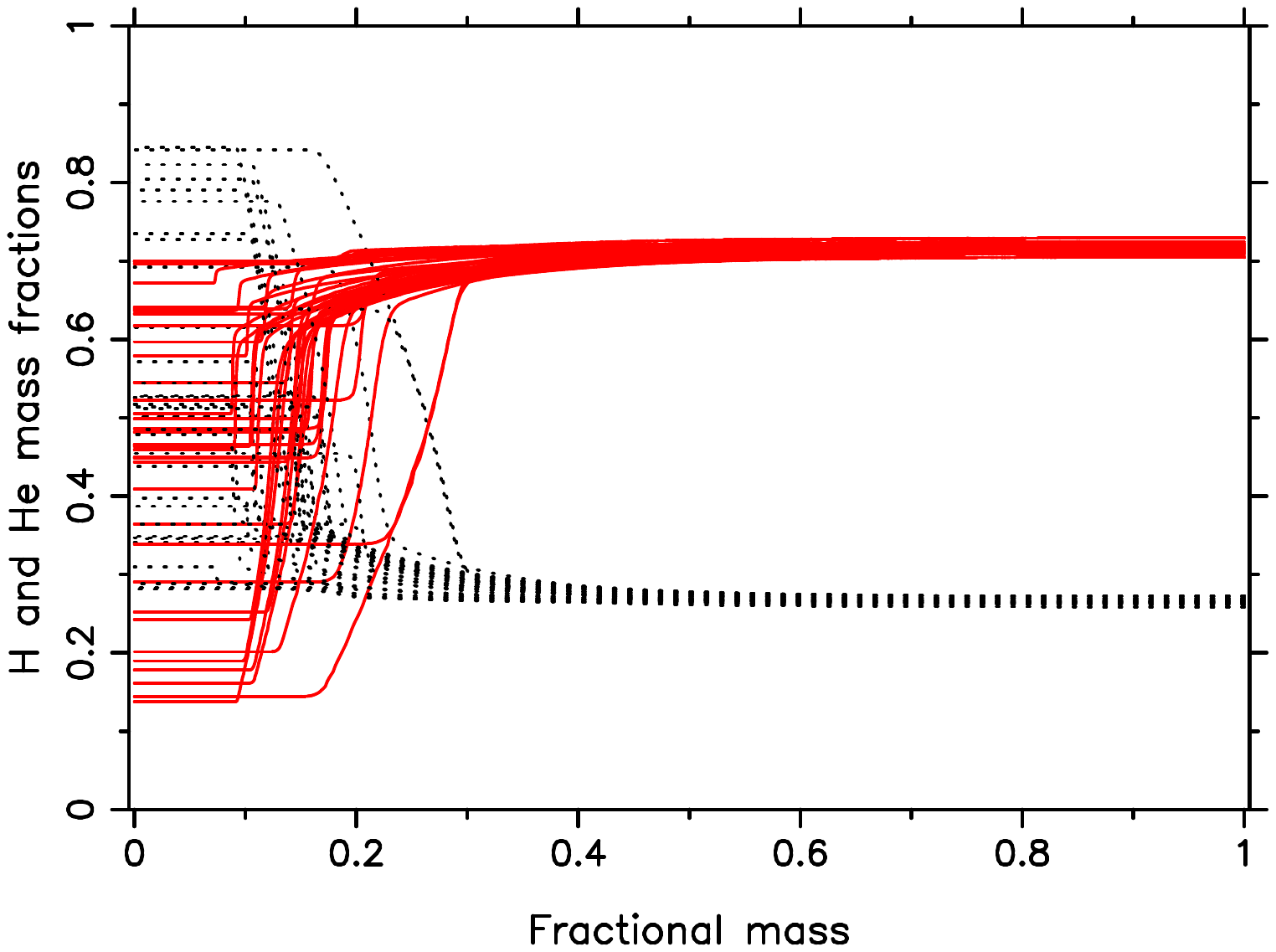}}}\hspace{0.2cm}
\rotatebox{270}{\resizebox{6.5cm}{!}{\includegraphics{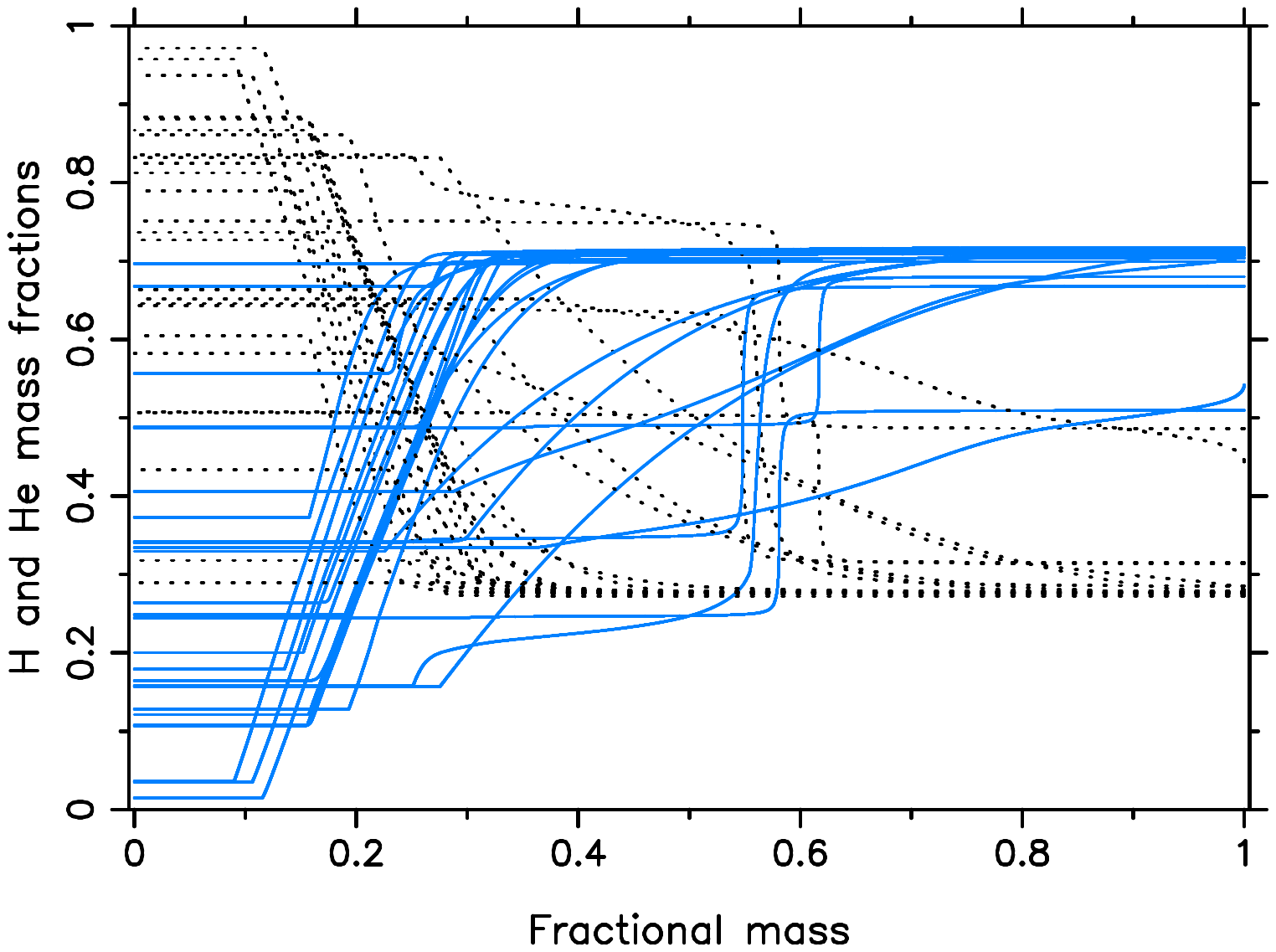}}}\\[0.2cm]
\rotatebox{270}{\resizebox{6.5cm}{!}{\includegraphics{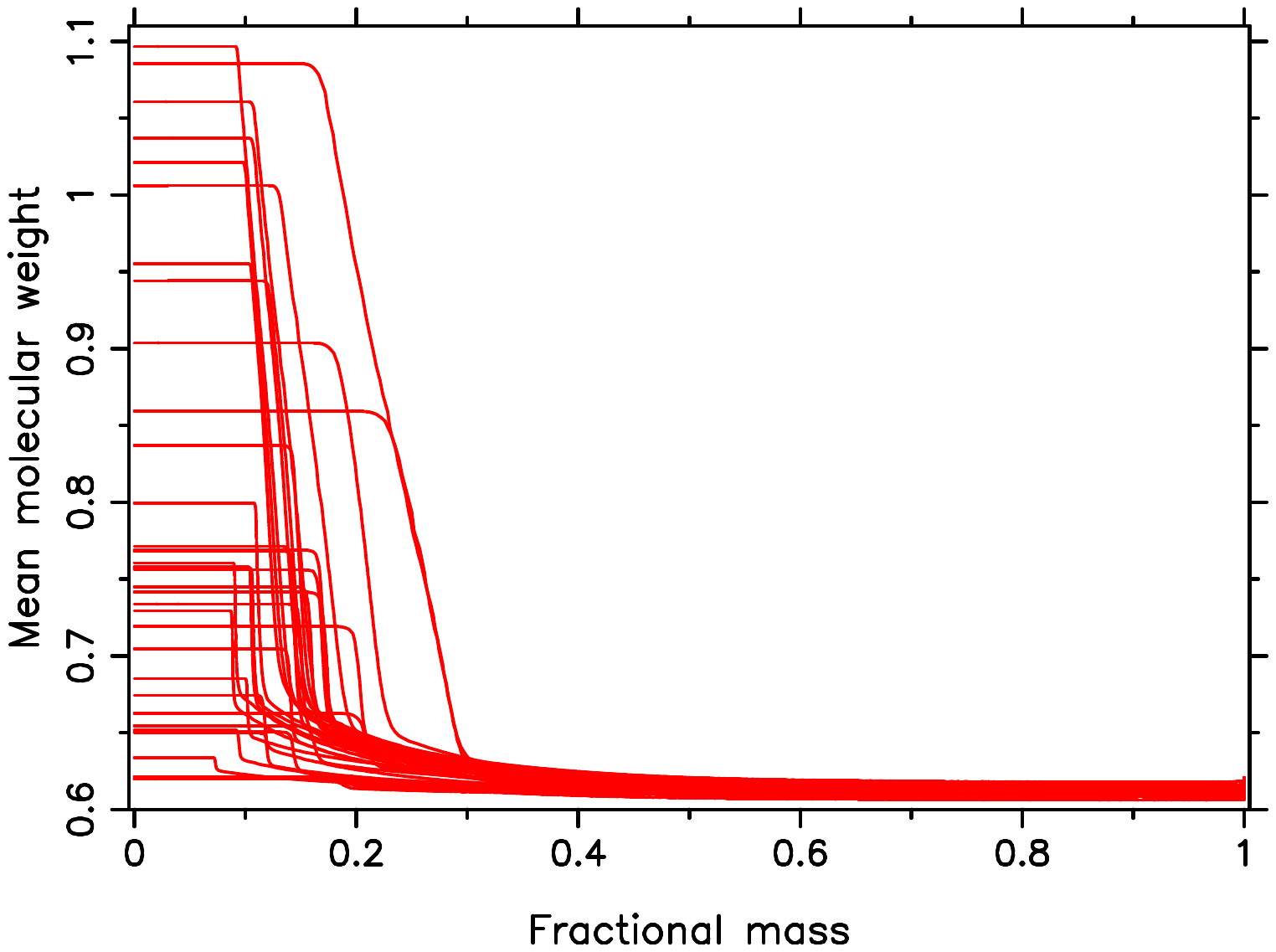}}}\hspace{0.2cm}
\rotatebox{270}{\resizebox{6.5cm}{!}{\includegraphics{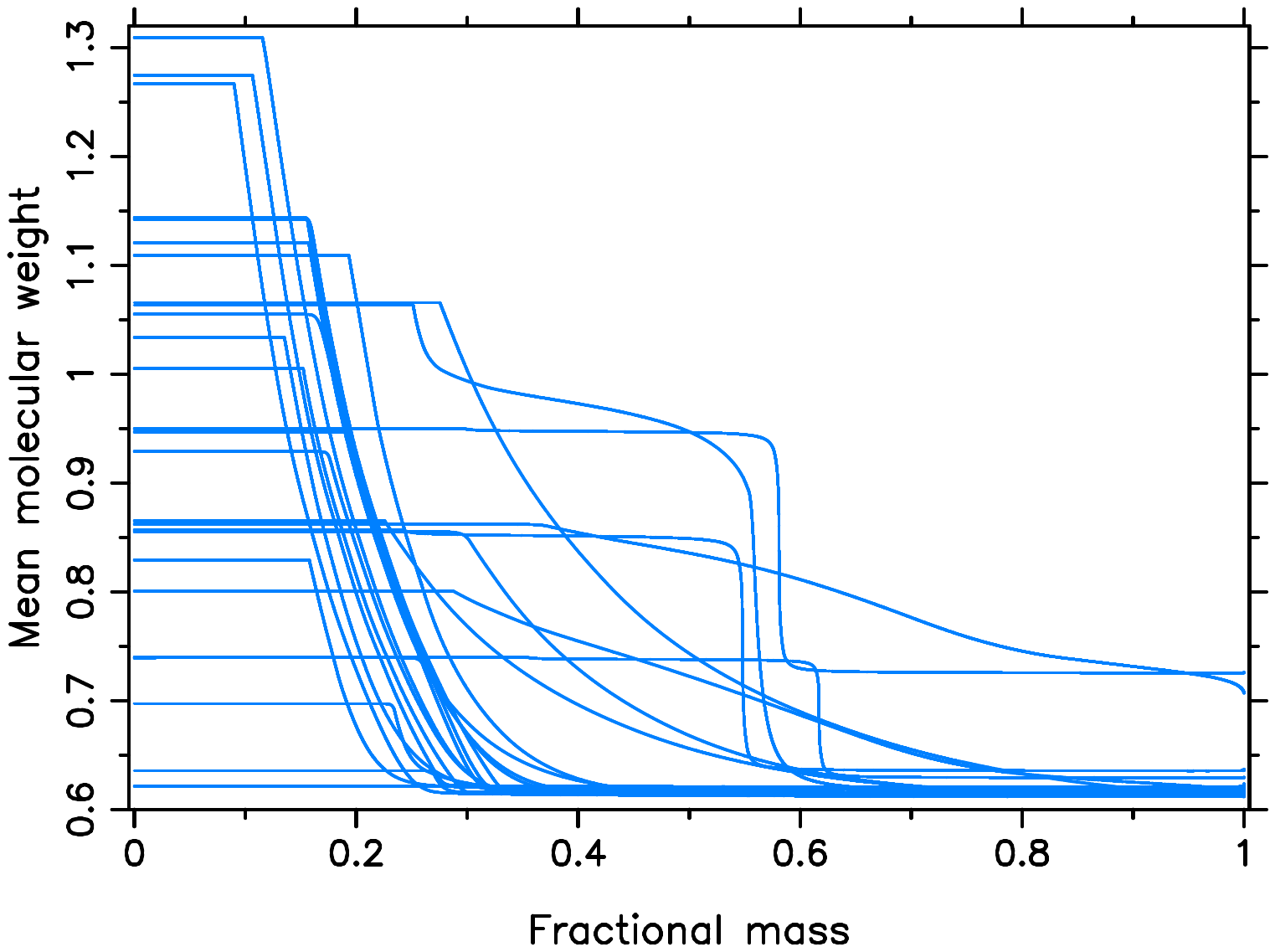}}}\\[0.2cm]
\rotatebox{270}{\resizebox{6.5cm}{!}{\includegraphics{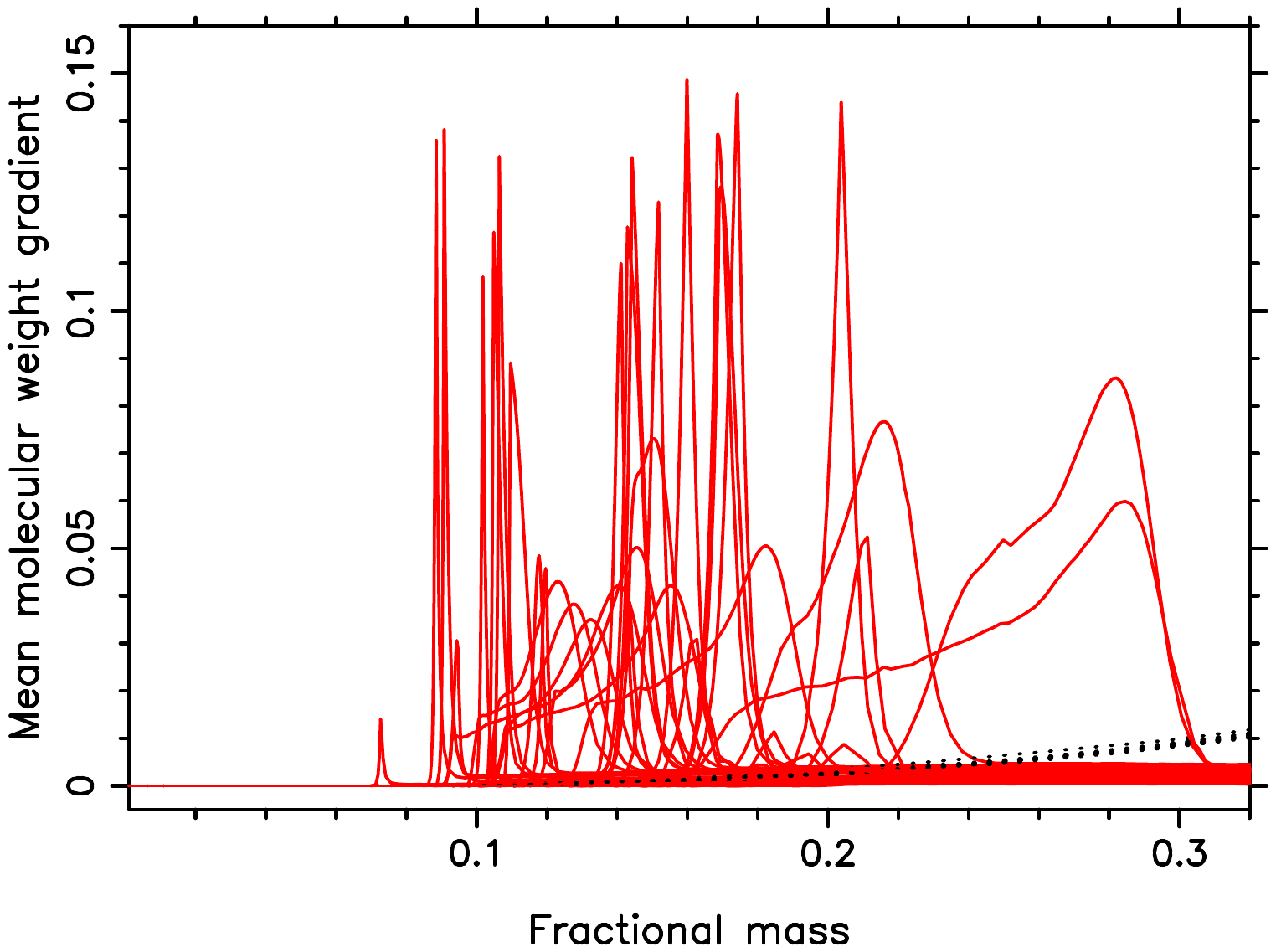}}}\hspace{0.2cm}
\rotatebox{270}{\resizebox{6.5cm}{!}{\includegraphics{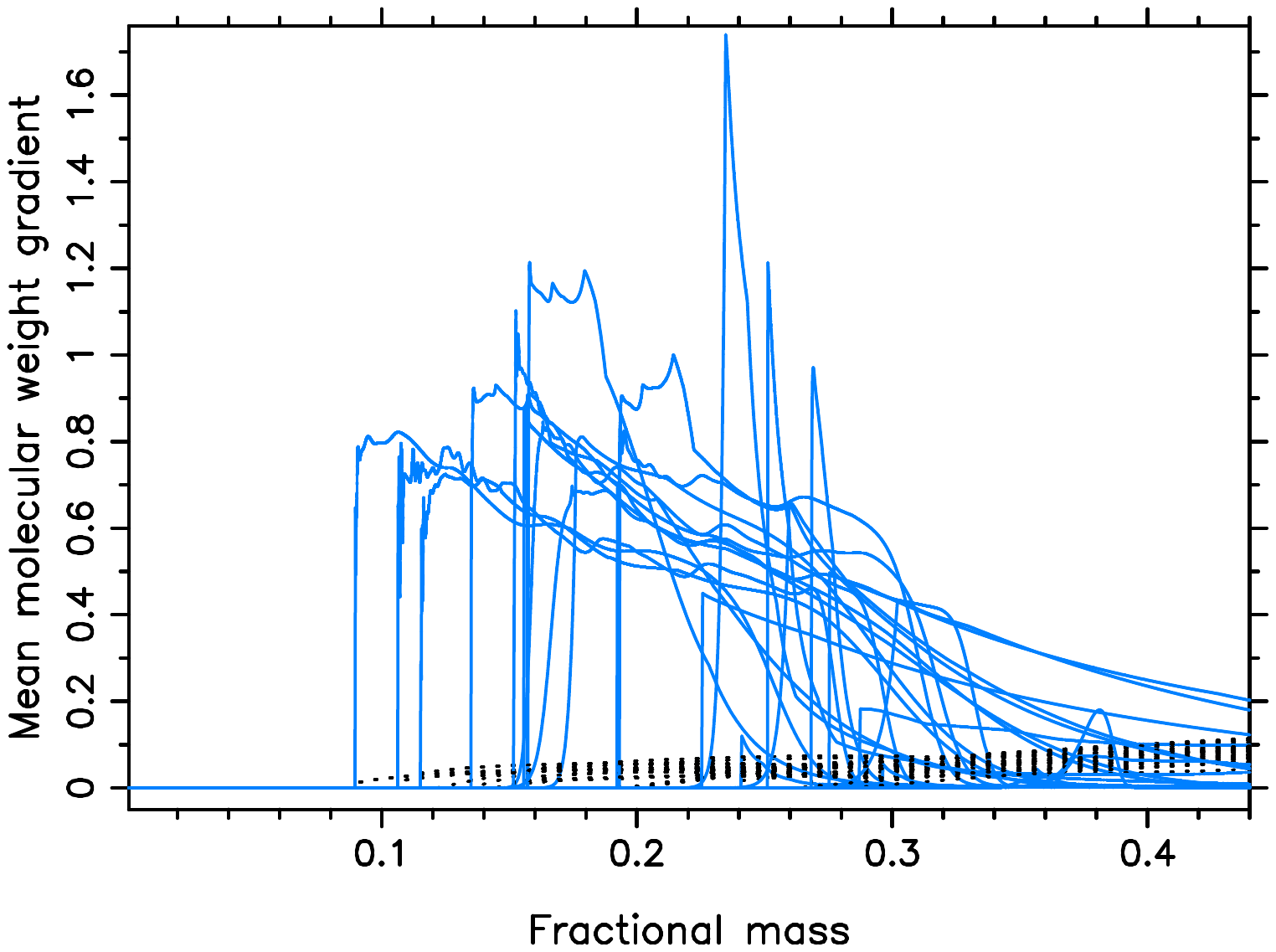}}}
\end{center}
\caption{\label{profiles} Profiles of hydrogen (full lines) and helium
  (dotted lines) mass fraction (top), of the mean molecular weight
  (middle), and of a zoom-in on its gradient in the area of the
  convective core (bottom) for the 37 $\gamma\,$Dor (left) and 26 SPB
  (right) stars. These profiles were retrieved from the best forward
  asteroseismic models computed by \citet{Mombarg2021} and by
  \citet{Pedersen2021}, respectively.}
\end{figure*}

We first recall the ingredients and definition of $\varepsilon$ for
the two extreme cases considered by \citet{Tokuno2022}, namely a
continuous profile of the hydrogen mass fraction at the convective
core boundary versus a discontinuity at that boundary resulting in a
sharp spike of the { BV} frequency $N(r)$.  Figure\,\ref{profiles}
shows the profiles of the hydrogen and helium mass fraction, mean
molecular weight ($\mu$) and its gradient ($\nabla_\mu$) for the 63
pulsators.  \citet{Aerts2021-GIW} already included the profiles for
$N(r)$ in their Fig.\,2. Here, we show $\nabla_\mu$, which provides by
far the largest contribution to $N(r)$ in the
{ boundary layer}
between the convective core and the radiative envelope. Indeed, many
of the $\gamma\,$Dor and all of the SPB stars experience a receding
convective core throughout their main sequence evolution, leaving
behind a gradient in $\mu$. This gradient is completely dominant over
the contribution of the entropy gradient shown as dotted lines in
the bottom panels of Fig.\,\ref{profiles}.  This figure reveals that
{ both smooth and abrupt profiles occur. Hence, we  used both
  formulations for the coupling coefficient
  deduced by \citet{Tokuno2022} and applied them to all  the stars.}
The upper two panels of  Fig.\,\ref{profiles}
show that the 26 SPB stars cover the entire main sequence,
while the sample of 37 $\gamma\,$Dor stars does not contain stars so
close to hydrogen exhaustion in the core.

Continuous hydrogen and helium mass fraction profiles at the core
boundary correspond to continuous density profiles. Assuming a constant
density in the convective core, which is a good approximation for the case of 
continuous profiles, \citet{Tokuno2022} derived the following expression
for the coupling coefficient $\varepsilon$ for a mode with frequency,
$\omega$\,:
\begin{equation}
  \label{eps1}
    \varepsilon \equiv\
  \displaystyle{
    \left(\frac{4\ \Omega^2}
         {r_a \cdot
\displaystyle{
           \left.
             \frac{{\rm d} N^2}{{\rm
                 d}r}\,\right|_{r=r_a}}}\ \right)^{\ 1/3},}
  \end{equation}
where $r_a$ is defined as the inner boundary of the envelope mode
propagation cavity where $\omega$ equals the { BV} frequency $N$.
This equation shows that the opportunity of mode coupling increases
with increasing rotation frequency and decreases with the steepness of
the wall caused by the stratification.  In practice $r_a$ is
{ close to}
the radius of the convective core (denoted here as $R_{\rm cc}$), which
is the position where $N^2$ becomes negative towards the stellar
centre. In the entire inner region where $N^2 < 0$, all the material
is homogeneously mixed, such that $\nabla_\mu=0$ (cf.\,bottom panels
of Fig.\,\ref{profiles}).  { However, $r_a$ is determined from the
  outer side towards the core, hence it depends on the properties
  of the boundary layer between the convective core and the
  envelope. The structure of this layer is still uncertain and
  asteroseismology has been used to try and unravel its chemical
  \citep{Pedersen2018} and temperature \citep{Michielsen2019}
  properties. Forward asteroseismic modeling of 26 SPB stars by
  \citet{Pedersen2021,Pedersen2022} pointed out that convective
  penetration with an adiabatic temperature gradient and full chemical
  mixing in the boundary layer is preferred above a radiative
  temperature gradient accompanied by a smooth diffusive mixing
  profile for somewhat over half of the stars.  Hence, this sample
  study of 26 pulsators showed that there is no unified best way to
  describe the temperature profile for all SPB pulsators.  For this
  reason, \citet{Michielsen2021} dissected the chemical mixing profile
  and the temperature gradient in the boundary layer of one of these 26 SPB
  pulsators in more detail. They compared models with a radiative
  temperature gradient with those having a gradual transition between
  an adiabatic and radiative gradient,
  where the transition is based on the P\'eclet number.
  The latter type of gradient is inspired by numerical simulations
  \citep{Viallet2013} and compares the ratio of advective
  versus diffusive transport in the boundarylayer between a
convection and radiative zone. 
\citet{Michielsen2021}  selected the SPB with the longest
  period spacing pattern to test their more detaild physical
  description of chemical mixing in the boundary layer,
  stitching a penetrative and diffusive mixing profile and assessing
  the nature of the temperature gradient. This
  turned out to be a challenging task, given the dominance of
  $\nabla_\mu$ over the temperature gradient in that layer
  (cf.\ Fig.\,\ref{profiles}). They found KIC\,7760680, which rotates
  at 25\% of its critical rate, to reveal a fully radiative rather than
  P\'eclet-based temperature gradient in their models with convective
  boundary mixing. Unraveling the temperature and mixing profiles for
  $\gamma\,$Dor pulsators is even harder than in the case of SPB stars
  \citep{Mombarg2019} because the effects of microscopic atomic
  diffusion cannot be ignored for these pulsators. Moreover, the
  role of radiative levitation for element mixing is unclear when
  treated without incorporating rotational mixing in the modeling
  \citep{Mombarg2020}. Although they are computationally intense, predictions
  of oscillation mode properties that take into account the joint effects
  of radiative levitation and rotation offer a promising route to
  bring the asteroseismic probing power of the boundary layer of
  $\gamma\,$Dor pulsators to the next level \citep{Mombarg2022}.}

For the case of discontinuous profiles of the hydrogen mass fraction
and of $N^2$ at the convective core boundary, \citet{Tokuno2022}
introduced an alternative expression for the coupling coefficient,
while also omitting  the assumption of a constant density, namely:
\begin{equation}
\label{eps2}
\displaystyle{
  \tilde{\varepsilon} \equiv
  \frac{\Omega}{N_0}\ {\rm with\ } N_0 \equiv \lim_{r\rightarrow R_{\rm cc}} N(r)},
    \end{equation}
where the limit has
to be taken from the radiative envelope towards the convective core,
that is, going from the envelope outside of the core towards the centre of the star. 

In the mathematical limit of ${\rm d}N^2/{\rm d}r \rightarrow\infty$,
one gets $\varepsilon \rightarrow 0$ preventing mode coupling {
  even} for the case of continuous profiles.
On the other hand, for strictly discontinuous profiles resulting in a
steep { BV} frequency spike, one would get $N_0 \rightarrow \infty$
and hence $\tilde{\varepsilon} \rightarrow 0$, again excluding mode
coupling. We evaluate the difference between the results from both
expressions in Eqs. \ref{eps1} and \ref{eps2} in the next section.

\subsection{Inferred values of $\varepsilon$ and  $\tilde{\varepsilon}$}

\begin{figure}
\begin{center}\vspace{-1cm} 
\rotatebox{270}{\resizebox{8.0cm}{!}{\includegraphics{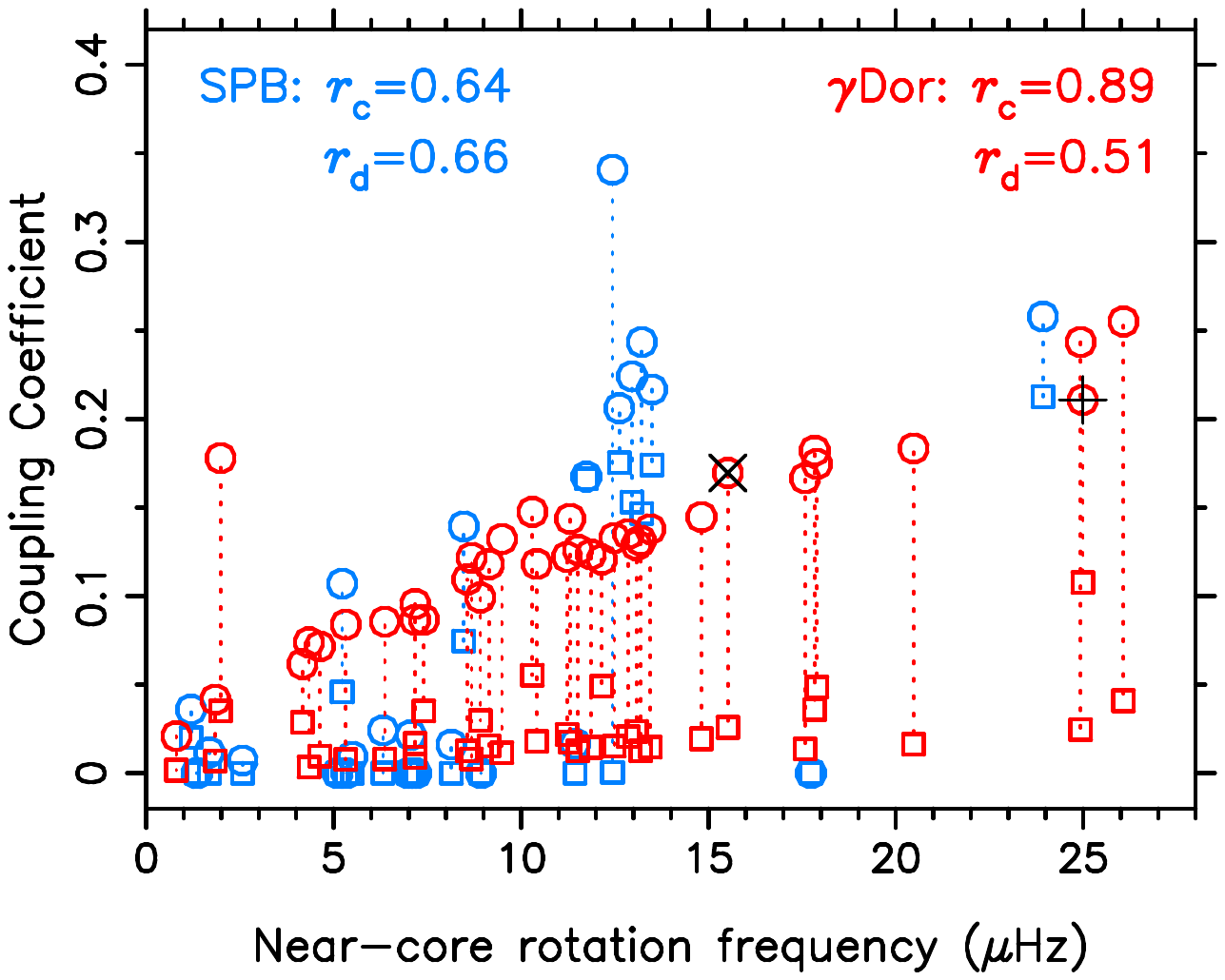}}}
\end{center}
\caption{\label{epsilon-omega} Values for $\varepsilon$ in the case of
  a continuous (circles) and for $\tilde{\varepsilon}$ 
  in the limit of a discontinuous (squares)
  { BV} profile plotted as a function of the near-core rotation
  frequency.  The $\gamma\,$Dor and SPB stars are indicated in red and
  blue, respectively. The two $\gamma\,$Dor stars for which
  \citet{Saio2021} detected a signature of coupling between a core and
  envelope mode are overplotted with black symbols, namely,
  KIC\,11907454 (cross) and KIC\,12066947 (plus).
  { Linear correlation coefficients, $r_{\rm c}$ and $r_{\rm d}$,
  between $\omega$ and $\varepsilon$, respectively
  $\tilde{\varepsilon}$,} are listed for the two samples. 
}
\end{figure}

The numerical values for $\varepsilon$ and $\tilde{\varepsilon}$ are
plotted as circles and squares, respectively, in Fig.\,\ref{epsilon-omega}
as a function of the near-core rotation frequency $\Omega$ taken from
\citet{Aerts2021-GIW}. First, we obtained values for
$\varepsilon$ and $\tilde{\varepsilon}$ between 0 and 1, as assumed by
\citet{Tokuno2022}. The numerical stellar evolution models computed by
\citet{Mombarg2021} and \citet{Pedersen2021} have a dense mesh for
their core boundary layers. This is necessary for proper computation
of the eigenmodes as they have high radial order and thus many nodes
need to get resolved for proper numerical approximation of the
displacement vector. These dense meshes give us good evaluation
capacity for the derivative in Eq.\,(\ref{eps1}) from its linearised
algebraic-differential version.  Since we here consider the general
case of inertial modes in the core with a range of frequencies
$\omega$, we compute the gradient from the value of $N^2$ and $r$ in
the three cells closest to the core boundary defined by $N^2(r) < 0$
and retain the largest value of $\varepsilon$ among the three, as a
good representation of maximal mode coupling { in the case of a
  continuous BV profile.}

The dotted lines in Fig.\,\ref{epsilon-omega} indicate the range
covered by $\varepsilon$ and $\tilde{\varepsilon}$ per star, connecting
the values obtained for the two approximations expressed by
Eqs. \ref{eps1} and \ref{eps2}, { which can be considered as
  representing the two extreme cases of reality}.
For some stars the two estimates are very similar
but for others the
difference between $\varepsilon$ and $\tilde{\varepsilon}$ is large.
This difference depends entirely on the shape of the $\nabla_{\mu}$
profile near the convective core boundary.

{ The strongest potential coupling between the
modes in the core and envelope systematically occurs for
$\varepsilon$ computed from Eq.\,(\ref{eps1}), as shown by comparing
the circles and squares per star in Fig.\,\ref{epsilon-omega}.} The
values we get for $\varepsilon$ are according to the theoretical
expectations in \citet{Tokuno2022}. We find that $\tilde{\varepsilon}$
is close to zero for many stars.
Low values for both approximations { of the coupling coefficient} occur for a
considerable fraction of the SPB pulsators, whose $\nabla_{\mu}$ value is
often some ten times larger than those for $\gamma\,$Dor stars, with a
sharp drop towards the convective core. A major conclusion thus is
that mode coupling is harder to establish for SPB than for
$\gamma\,$Dor stars. { As discussed in the previous section, this}
is in agreement with {\it Kepler\/} space
photometry, where mode coupling between core and envelope is not yet
reported for any of the SPB stars, while it was found in various of
the $\gamma\,$Dor stars and interpreted as such by
\citet{Saio2021}. We refer to the latter paper for an extensive
discussion and illustrations of the delicate balance between the local
rotation and two pulsation frequencies of the two involved modes and
their eigenmode properties for the mode coupling to become effective.

\citet{Saio2021} found two
of the $\gamma\,$Dor stars in our sample to have
a dip in the period spacing pattern that is characteristic of the coupling
between a core inertial mode and an envelope gravito-inertial mode,
namely, KIC\,11907454 and KIC\,12066947. These have a relatively high
$\varepsilon$ value and are among the most rapid
rotators, as indicated in Fig.\,\ref{epsilon-omega}.

The lack of any mode coupling detection for SPB stars may simply be
due to an observational bias, as there are a factor 20 less such stars
with period spacing patterns from space photometry compared to
$\gamma\,$Dor stars.  The SPB star KIC\,8255796 may be the best
candidate, as it has a typical Lorentz-shape dip in its period spacing
pattern \citep[][see Fig.\,15 in the supplementary
  material]{Pedersen2021}.  However, its pattern is among the shortest
ones, constituting of only nine identified dipole modes. This SPB has a
mass of 5.7\,M$_\odot$ and a relatively slowly rotating near-core
boundary layer (some 19\% of the star's critical rate). Moreover, it
has a low $\varepsilon=0.0118$ and is the most evolved of all the 26
SPB stars in the sample while, as we will show below, we expect the
coupling capacity to be weakest in that stage of the main sequence.
Therefore, we consider it unlikely for its observed dip structure to be
due to mode coupling.

Since we wish to { understand how mode coupling between core and
  envelope modes comes about, we now investigate relationships between
  $\varepsilon$ and $\tilde{\varepsilon}$ with respect to other
  parameters characterizing the convective core.  Following their
  definitions in Eqs. \ref{eps1} and \ref{eps2}, a positive
  correlation occurs with $\Omega$; namely, higher rotation rates lead
  to higher coupling coefficients.  This is indeed found for both
  $\varepsilon$ and $\tilde{\varepsilon}$ (shown in
  Fig.\,\ref{epsilon-omega}), where the linear correlation coefficients
  for the continuous and discontinuous cases are denoted as $r_{\rm
    c}$ and $r_{\rm d}$, respectively. They are almost equal ($r_{\rm
    c}=0.64$ and $r_{\rm d}=0.66$) for the SPB pulsators, while the relation is
  more pronounced for the continuous case of the $\gamma\,$Dor star models
  ($r_{\rm c}=0.89$ versus $r_{\rm d}=0.51$).}

\subsection{Relation with stellar core parameters}

\begin{figure}
\begin{center}\vspace{-1cm} 
\rotatebox{270}{\resizebox{8.0cm}{!}{\includegraphics{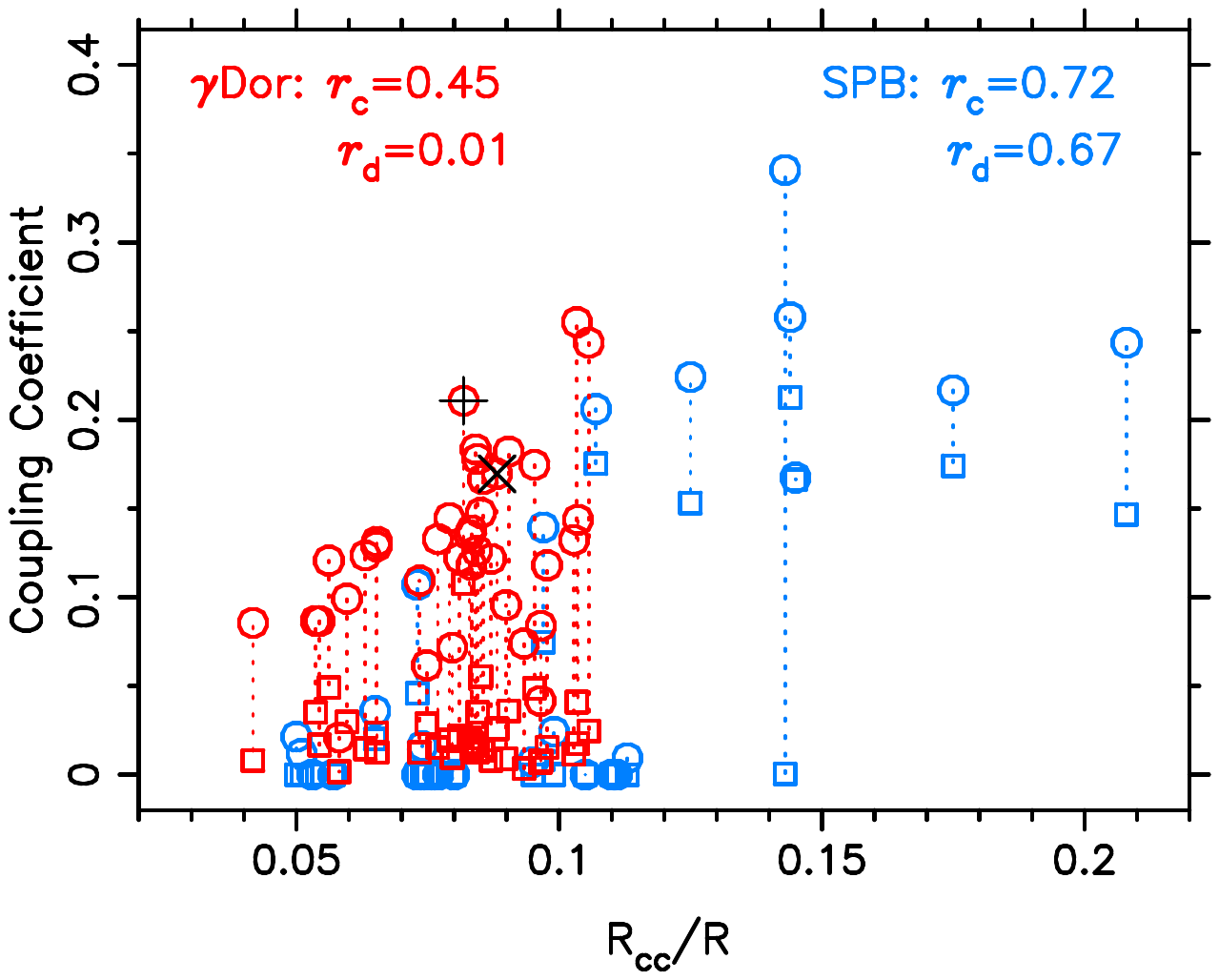}}}
\end{center}
\caption{\label{epsilon-Rcc} Same as Fig.\,\ref{epsilon-omega}, but
  plotted as a function of the extent of the convective core, expressed
  as a fraction of the total radius.}
\end{figure}

Figure\,\ref{epsilon-Rcc} shows a fairly strong positive correlation
between
$\varepsilon$ { or $\tilde{\varepsilon}$}
  and the relative size of the convective core for
  SPB stars. The trend also occurs for { $\varepsilon$ of}
  $\gamma\,$Dor stars { but
  with a lower $r_{\rm c}$ value -- yet it is absent for $\tilde{\varepsilon}$.}
This { difference} can be understood from 
the properties of the fractional radius among the two
samples. Indeed, the $\gamma\,$Dor stars cover a narrower range in
$R_{\rm cc}/R$ (with $R$ the radius of the star) than the SPB stars,
implying a weaker correlation between $\varepsilon$ and convective
core size than for the SPB stars.  Moreover, $\gamma\,$Dor stars cover
a range in stellar mass ($M$) such that some stars in the sample
exhibit a growing convective core { with a steep BV profile} during the core hydrogen burning,
while others have a receding core { and a smoother varying BV frequency}.
This mixed behaviour in terms of
convective core size evolution makes it intrinsically less obvious
to have a tight correlation between $\varepsilon$ and $R_{\rm cc}$
{ and explains the absence of any relation with $\tilde{\varepsilon}$.}
The two $\gamma\,$Dor stars with detected mode coupling do not stand
out in terms of their $R_{\rm cc}/R$ value.

\begin{figure}
\begin{center}\vspace{-1cm} 
\rotatebox{270}{\resizebox{8.0cm}{!}{\includegraphics{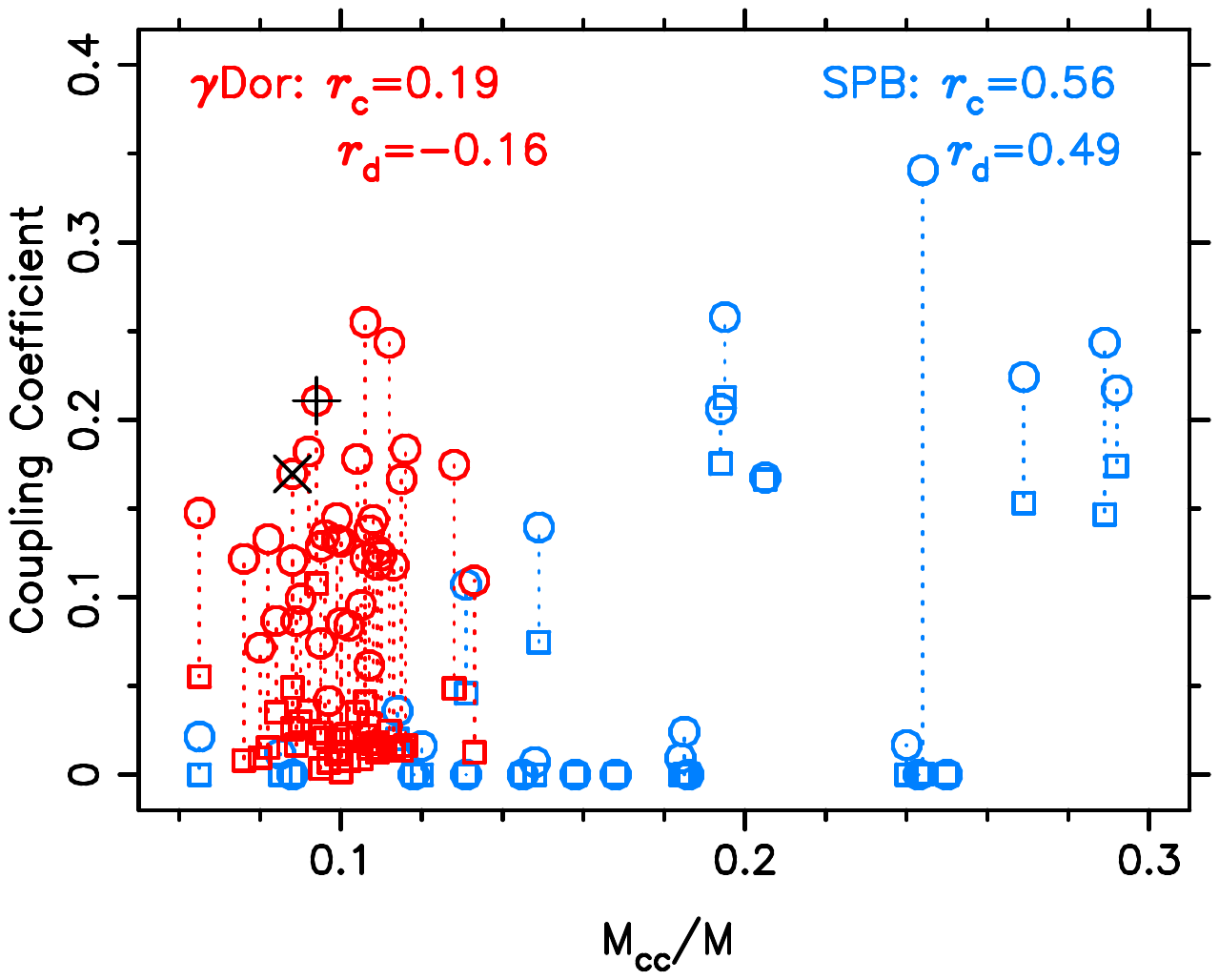}}}
\end{center}
\caption{\label{epsilon-Mcc} Same as Fig.\,\ref{epsilon-omega}, but
  plotted as a function of
 the mass in the convective core, expressed
  as a fraction of the total stellar mass.}
\end{figure}

For both samples, the convective core mass (expressed as relative
fraction of the total stellar mass) has a lower linear correlation
with respect to $\varepsilon$ than the fractional radius of the
convective core { and the correlation essentially disappears for
  the $\gamma\,$Dor stars} 
(see Fig.\,\ref{epsilon-Mcc}). Again, this is to be
expected as the $\gamma\,$Dor sample covers stars with a
growing and decreasing convective core mass.
For both samples $r_{\rm  c}$  is
lower for $M_{\rm cc}$ than for $R_{\rm cc}$ because the amount of
hydrogen that gets injected into the core at the expense of CNO
products is more dependent on the cause and character of the overshoot
and envelope mixing than the size of the core. We also checked for
correlations between $\varepsilon$ and the values of $R_{\rm cc}$ or
$M_{\rm cc}$ themselves instead of their fractional dimensionless
values, but this leads to { even lower $r_{\rm c}$ and $r_{\rm d}$ values}
  than those listed in
Figs.\,\ref{epsilon-Rcc} and \ref{epsilon-Mcc}.

\begin{figure}
\begin{center}\vspace{-1cm} 
\rotatebox{270}{\resizebox{8.0cm}{!}{\includegraphics{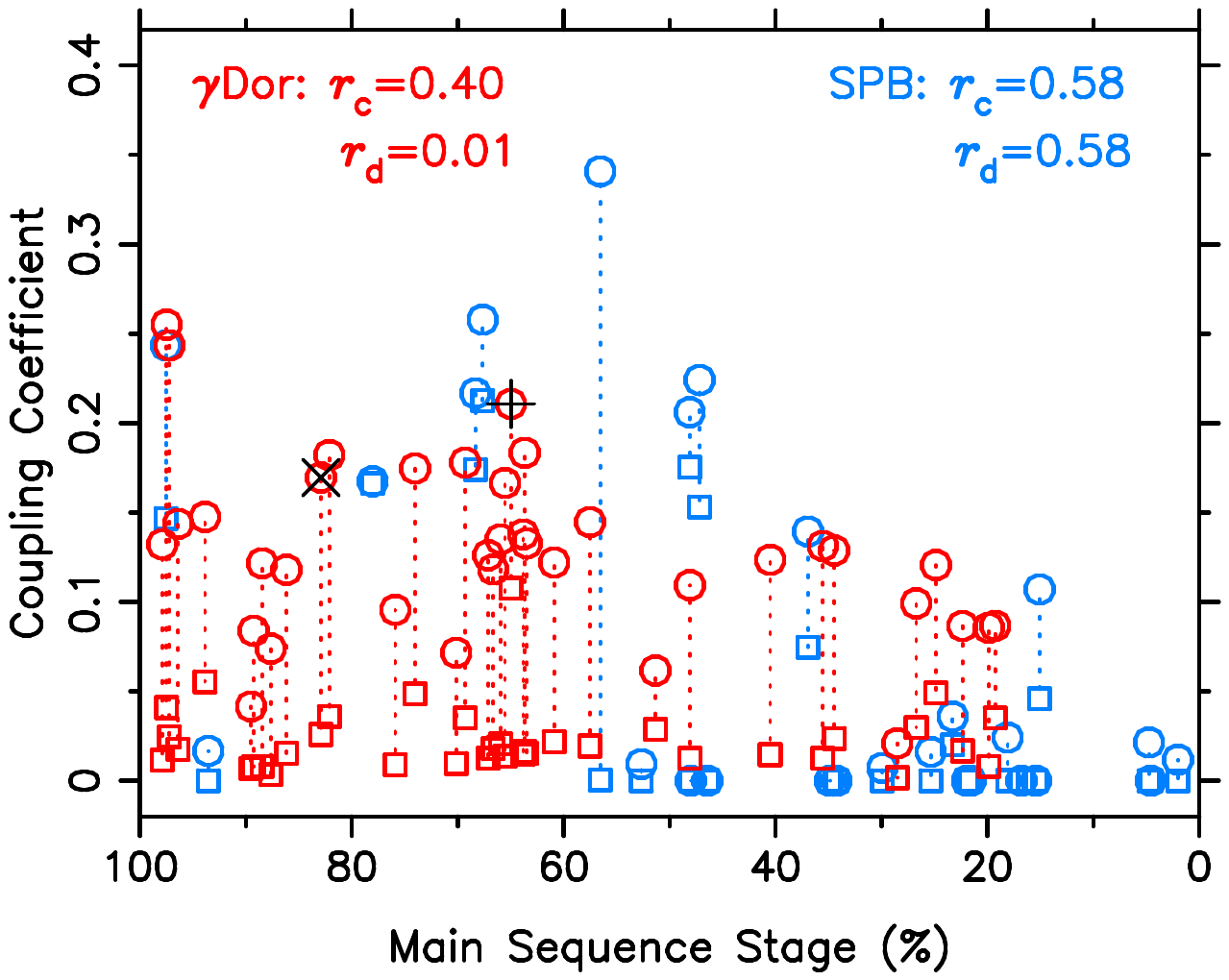}}}
\end{center}
\caption{\label{epsilon-Xc} Same as Fig.\,\ref{epsilon-omega}, but
  plotted as a function of the main sequence stage, defined as the
  current core hydrogen mass fraction divided by the initial hydrogen
  mass fraction at birth.}
\end{figure}

Both $R_{\rm cc}/R$ and $M_{\rm cc}/M$ evolve as the core hydrogen
burning progresses. The manner in which they do depends greatly on the level and
character of the internal mixing in the core boundary layer and in the
envelope
\citep{Moravveji2015,Moravveji2016,Mombarg2019,Mombarg2021,Pedersen2021,Pedersen2022}. Thus, a
comparison between $\varepsilon$ { or $\tilde{\varepsilon}$} and
the evolutionary stage is also of
interest. Figure\,\ref{epsilon-Xc} reveals the { relationships
  between either $\varepsilon$ or $\tilde{\varepsilon}$} and the
central core hydrogen mass fraction, $X_c$, with respect to the
initial value at birth.  This fraction represents the stage of the
main sequence evolution.  We find a relatively mild trend that mode
coupling { in the continuous BV case} is more likely in the early
stages of stellar evolution. This trend is similar to the one found
for $R_{\rm cc}/R$, that is somewhat tighter for the SPB stars than
for the $\gamma\,$Dor stars. This is in line with the fact that the
convective core of SPB stars decreases monotonically from birth to
central hydrogen exhaustion for the entire sample, while this is not
the case for the $\gamma\,$Dor sample. { No connection was found
  between $\tilde{\varepsilon}$ and the evolutionary stage for
  $\gamma\,$Dor stars, while $\varepsilon$ and $\tilde{\varepsilon}$
    behave similarly with respect to the evolutionary stage for the SPB class.}

\begin{figure}
\begin{center}\vspace{-1cm} 
\rotatebox{270}{\resizebox{8.0cm}{!}{\includegraphics{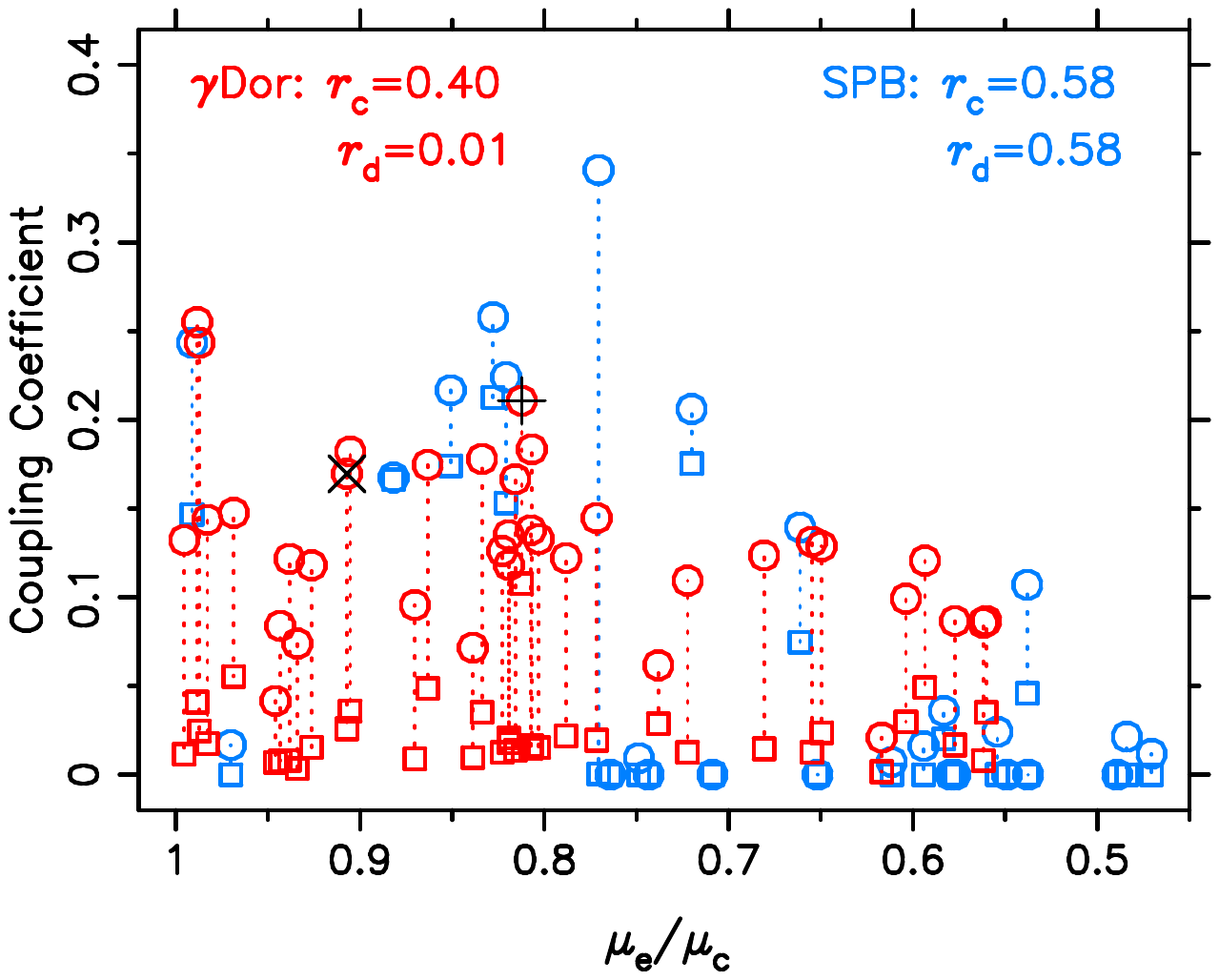}}}
\end{center}
\caption{\label{epsilon-mu} Same as Fig.\,\ref{epsilon-omega}, but
  plotted as a function of $\mu_{\rm e}/\mu_{\rm c}$.}
\end{figure}

The chemical evolution of the star is roughly captured by the value of
the mean molecular weight in the envelope versus that in the core,
$\mu_{\rm e}/\mu_{\rm c}$. This ratio is 1 at the star's birth and
decreases as the chemical evolution of the star continues.  However,
in contrast to the evolutionary stage shown in Fig.\ref{epsilon-Xc},
$\mu_{\rm e}/\mu_{\rm c}$ depends directly on the effect of internal
mixing in the radiative envelope.  We show the relationship between
$\varepsilon$ { or $\tilde{\varepsilon}$}
and $\mu_{\rm e}/\mu_{\rm c}$ in Fig.\,\ref{epsilon-mu}.
This gives the same picture as the one seen for the main sequence stage.

\subsection{On the Sch\"onberg-Chandrasekhar limiting mass}

A similar yet slightly different way of looking at the evolutionary
aspect of the mode coupling coefficients
is via the so-called Sch\"onberg-Chandrasekhar limiting mass,
$M_{\rm SC}$ \citep{SchCh1942}.  This quantity is formally defined as the
maximum mass a helium core can have after hydrogen exhaustion in the
core in order to remain inert, that is to withstand the pressure of
the encompassing stellar envelope without contracting.  When the value
of $M_{\rm SC}$ is surpassed, the helium core will start to shrink and
the star will evolve on a fast contraction time scale rather than on a
slow nuclear time scale.

\begin{figure}
\begin{center}\vspace{-1cm} 
\rotatebox{270}{\resizebox{8.0cm}{!}{\includegraphics{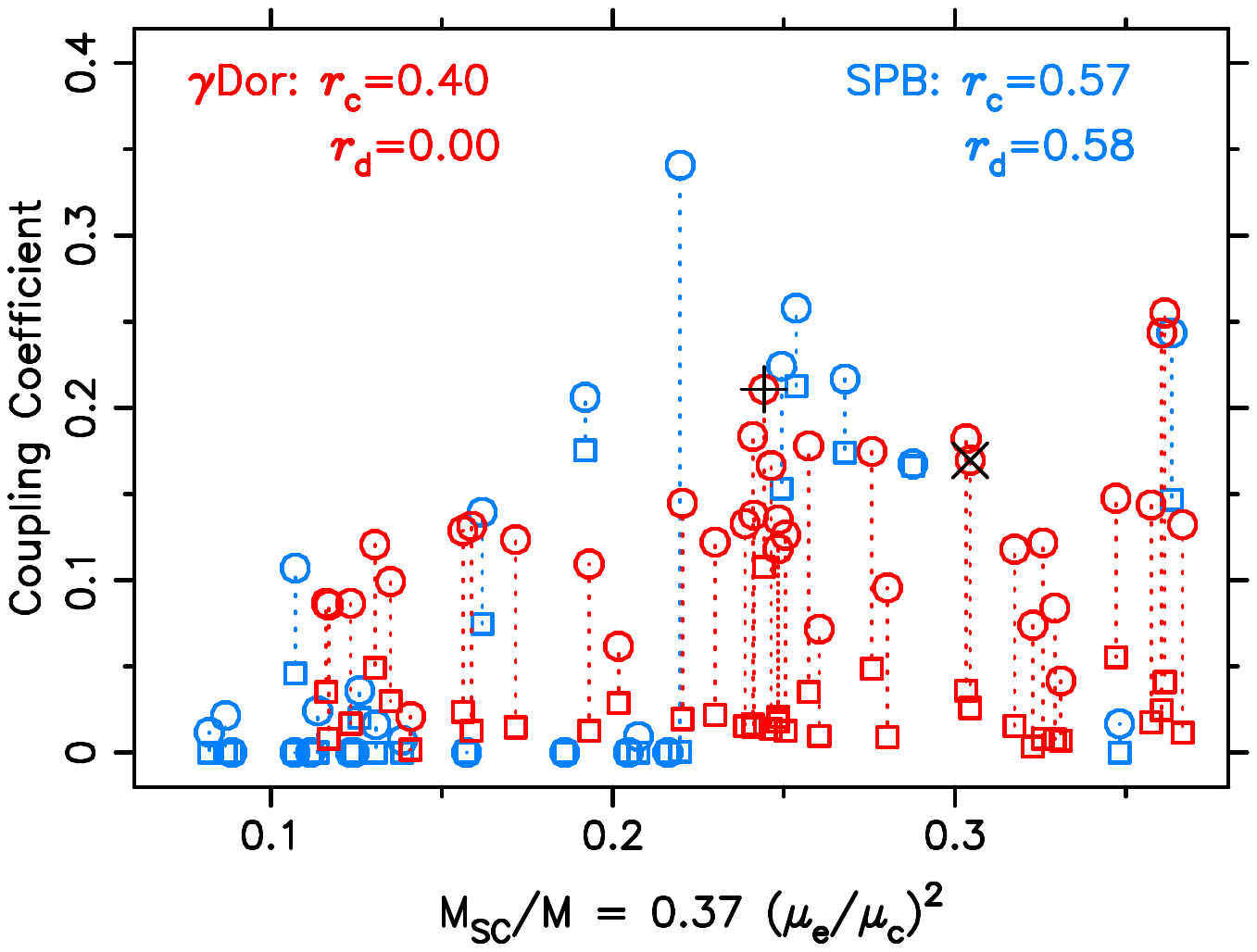}}}
\end{center}
\caption{\label{epsilon-Msc} Same as Fig.\,\ref{epsilon-omega}, but
  plotted as a function of the Sch\"onberg-Chandrasekhar core mass
  limit. }
\end{figure}

An analytical expression for $M_{\rm SC}$ has been deduced from the
virial theorem in the case of non-rotating polytropic stellar models
with an isothermal helium core \citep{Stein1966,CoxGiuli1968}. Despite
the fact that real stars do not adhere to a polytropic equation of
state, the approximation for the Sch\"onberg-Chandrasekhar limit
deduced from numerical stellar models was found to be essentially the
same as for the polytropic approximation \citep{Kippenhahn2013}:
\begin{equation}
\label{SchCh}
  M_{\rm SC}/M \simeq 0.37
\cdot \left( \mu_{\rm e}/\mu_{\rm c} \right)^2
,\end{equation}
 \citet{Maeder1971} investigated the effect of
uniform rotation on $M_{\rm SC}$, concluding that its value does not
change appreciably. It decreases due to rotation by at most 3\%, even
for fast rotation close to the critical value. It is thus meaningful
to compute the current value of $M_{\rm SC}/M$ from Eq.\,\ref{SchCh}
for our sample of rapidly rotating gravito-inertial mode pulsators,
some of which are close to hydrogen exhaustion in their core
(Fig.\,\ref{epsilon-Xc}).  Figure\,\ref{epsilon-Msc} shows the
relation between $\varepsilon$ { or $\tilde{\varepsilon}$} 
and the quadratic dependence on
$\mu_{\rm e}/\mu_{\rm c}$ via the computed value of $M_{\rm SC}$ from
the asteroseismically calibrated $\mu$ profiles shown in
Fig.\,\ref{profiles}.

The correlation between $M_{\rm SC}/M$ and $\varepsilon$ plotted in
Fig.\,\ref{epsilon-Msc} is obviously consistent with the one displayed
for the linear dependence, $\mu_{\rm e}/\mu_{\rm c}$, in
Fig.\,\ref{epsilon-mu}.  The obtained numerical values for $M_{\rm
  SC}/M$ in the current evolutionary stage of the stars computed from
Eq.\,\ref{SchCh} resemble those of the fractional convective core
mass used in Fig.\,\ref{epsilon-Mcc}.  The expression in
Eq.\,\ref{SchCh} was deduced for standard stellar models at core
hydrogen exhaustion, irrespective of the kind and level of mixing that
the star underwent during the main sequence.  While
\citet{Maeder1971} found that rotation hardly affects it, the chemical
evolution of the star due to near-core boundary mixing
\citep{Michielsen2021} and envelope mixing \citep{Pedersen2021} does
play a role in the behaviour of $\mu$. We therefore compare the
asteroseismically inferred values for $M_{\rm cc}/M$ at the current
evolutionary stage of the stars with the current value of $M_{\rm
  SC}/M$ computed via Eq.\,\ref{SchCh}.

\begin{figure}
\begin{center}\vspace{-1cm} 
\rotatebox{270}{\resizebox{8.0cm}{!}{\includegraphics{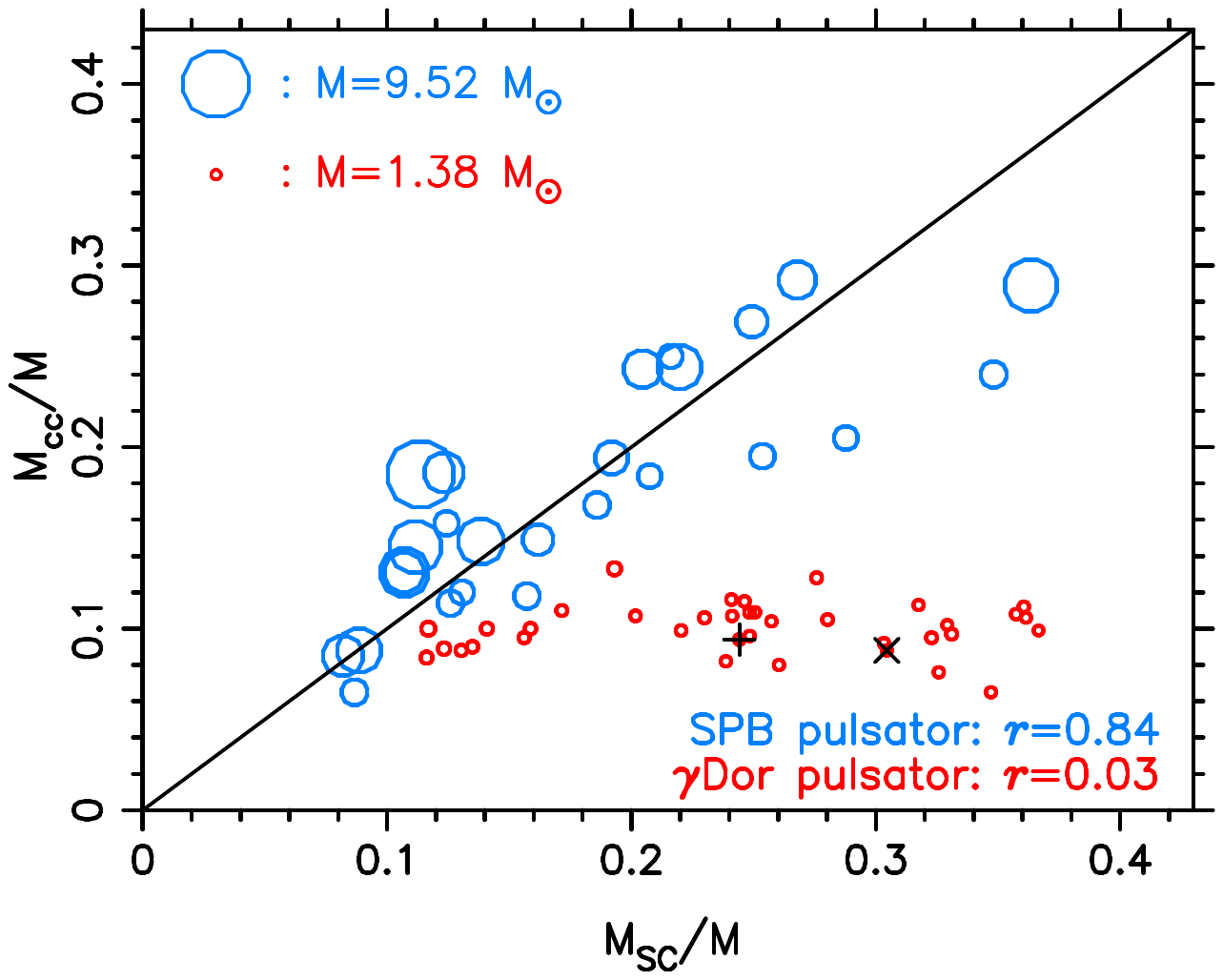}}}
\end{center}
\caption{\label{Mcc-Msc} Convective core mass versus
  Sch\"onberg-Chandrasekhar mass limit expressed as a fraction of the
  total mass. The $\gamma\,$Dor and SPB stars are indicated in red and
  blue symbols, respectively. The symbol size scales
  linearly with the total stellar mass, within the two extreme masses
  occurring in the samples, as indicated in the legend.}
\end{figure}

Our sample contains stars with masses between 1.38\,M$_\odot$ and
9.52\,M$_\odot$. This lower limit is close to the value where stars
transition from being below to above their $M_{\rm SC}$ limit at
the hydrogen exhaustion.  As long as the helium core mass remains
below $M_{\rm SC}$, the core can stay inert at hydrogen exhaustion
and the subsequent hydrogen shell burning happens on a
nuclear timescale. If, in contrast, the helium core mass
exceeds $M_{\rm SC}$, it will start to contract while hydrogen burns
in a shell and so this stage of evolution happens on a much shorter
contraction time scale. For this reason, the value of $M_{\rm SC}$
is an important quantity for the star's evolution.
We compare the current values of
$M_{\rm cc}/M$ and $M_{\rm SC}/M$ according to Eq.\,\ref{SchCh} for our
63 sample stars in Fig.\,\ref{Mcc-Msc}, where the symbols are linearly
scaled with the total stellar mass. During their evolution, stars
evolve from the right to the left and those with a shrinking
convective core also from the top to the bottom in such a diagram.  It
can be seen that all of the $\gamma\,$Dor stars have a convective core
mass unrelated to and below their Sch\"onberg-Chandrasekhar
limit. They will steadily evolve almost horizontally to the left in
the figure as they approach hydrogen depletion, because they hardly
experience envelope mixing and thus keep their $\mu_{\rm e}$
unchanged.  Since none of them are close to hydrogen exhaustion and
given their mass range, it is not expected  that they would have already surpassed
their $M_{\rm SC }$ value.

The more massive SPB stars, on the other hand, have convective cores
tightly correlated with the evolution of their $M_{\rm SC}/M$ limit,
already surpassing that limit on the main sequence for the more
massive sample stars as expected.  For completeness, we also used the
new fourth-degree polynomial fit for $M_{\rm SC}$ proposed by
\citet[][Eq.\,(15) in their manuscript]{Chowdhury2023} instead of the second-degree formula in Eq.\,(\ref{SchCh}). This gives the same
  results as those in Fig.\,\ref{Mcc-Msc} for the linear correlation
  coefficients of { $r_{\rm c}$ and $r_{\rm d}$}
    and for the plot to remain within the symbol sizes.

\citet{Pedersen2022} predicted the values of the helium core masses at
core hydrogen exhaustion of the 26 SPB stars based on their current
asteroseismic modes and $M_{\rm cc}$ values. She concluded that due
to their levels of envelope mixing, most of these SPB stars will have
higher helium core masses than those resulting from standard stellar
evolution without extra mixing. Her results are in line with
convective core mass estimations from massive eclipsing binaries
\citep{Tkachenko2020,Johnston2021}. These binary and asteroseismic
results for higher-than-standard convective core masses have not yet
been taken into account in chemical yield computations guiding the
overall chemical evolution models of galaxies
\citep{Karakas2014,Kobayashi2020}. However, the  higher levels of
core masses that have been measured will have a major impact on yield prediction models, given
that the uncertainties of such models for intermediate-mass stars
mainly come from unknown internal mixing.

\section{Discussion and future prospects}

In this paper, we provide numerical values for the coupling
{
coefficients,
$\varepsilon$ and $\tilde{\varepsilon}$,
which are unit-less measures below value 1}
indicative of the opportunity to have inertial modes in rapidly
rotating convective cores couple to gravito-inertial envelope
modes. We deduced numerical values for $\varepsilon$ { and
  $\tilde{\varepsilon}$} from forward asteroseismic models of 63
  gravity-mode pulsators, covering a mass range from 1.38\,M$_\odot$ to
  9.52\,M$_\odot$. Our findings are in agreement with the theoretical
  expectations and interpretations recently proposed by
  \citet{Tokuno2022}. From the perspective of the $\varepsilon$ {
    or $\tilde{\varepsilon}$ values with respect to the near-core
    rotation frequency}, the opportunity for mode coupling is similar
  for SPB and $\gamma\,$Dor star models. In practice, however, the
  sample of the few tens of SPB stars with identified modes from
  period spacing patterns available in the literature contains slower
  rotators than many of the $\gamma\,$Dor stars in the   sample of published pulsators with proper diagnostic gravity-mode
  patterns, which is 20 times larger \citep[see ][for an overview]{Aerts2021}.

  We find that the inferred values of $\varepsilon$
{ or $\tilde{\varepsilon}$}
  offer a useful
diagnostic to hunt for core-to-envelope mode coupling. Yet
the
{ values of $\varepsilon$ and $\tilde{\varepsilon}$
  alone do} not allow us to distinguish among the
few gravito-inertial pulsators with a core mode coupled to a
gravito-inertial mode from the majority of them that do not reveal
this coupling phenomenon. The values of $\varepsilon$
{ and $\tilde{\varepsilon}$}
for the two
$\gamma\,$Dor stars included in our sample of 37 { do} not stand out
from the other rapid rotators in terms of their core properties
deduced from forward modeling of their identified gravito-inertial
modes.  A visual inspection of characteristic dips in the period spacing
patterns as done by \citet{Saio2021} thus remains the best (and
currently only) way to find pulsators with coupled inertial core
modes. These selected targets then allow for the derivation of the
rotation frequency in their convective core via the matching of the
frequencies of the inertial and gravito-inertial modes, given the
near-core rotation frequency measured from the tilt of their period
spacing pattern(s). Such type of core rotation measurement was first
proposed by \citet{Ouazzani2020} and further elaborated upon by
\citet{Saio2021}. This is currently the best way to constrain the
internal rotation profile from gravity-mode pulsators, given the
challenges encountered for frequency inversions for this type of
pulsators caused by nonlinear effects, notably the dense occurrence of
avoided crossings \citep{Vanlaer2023}.

The observational challenge remains to distinguish dips in the period
spacing patterns stemming from inertial mode coupling rather than from
periodic deviations induced by mode trapping at the bottom of the
envelope due to a strong $\mu$-gradient in that position, because both
phenomena occur simultaneously. This also explains the current absence
of any mode coupling detection in SPB stars so far, given their much
larger $\nabla_{\mu}$ value and more extended $\mu-$gradient zone
in the near-core boundary layer compared to the one
of $\gamma\,$Dor stars (cf.\,Fig.\,\ref{profiles}). On the other
hand, this observational challenge is in agreement with the population
statistics. Indeed, we currently have about 28 SPB stars with period
spacing patterns and none of them have a convincing dip structure that would be expected for inertial
mode coupling. \citet{Saio2021} found 16 of the current sample of 670
$\gamma\,$Dor stars to have the signature (2.4\%). While it
concerns a low number of stars, these are crucial targets to map the
internal rotation
\citep[and possibly the magnetic field, see e.g.\,][]{VanBeeck2020,Dhouib2022}
of intermediate- and
high-mass stars. In this respect, the potential of the TESS mission
has yet to be explored. Indeed, the 61 TESS $\gamma\,$Dor and 2 SPB
stars with detected period spacing patterns found by
\citet{Garcia2022b,Garcia2022a} are promising in this respect, but
their period-spacing patterns from 352\,d light curves are too short to
offer proper dip structures caused by mode coupling or mode
trapping. However, progress can be ensured by analyzing the thousands of
light curves for the SPB and $\gamma\,$Dor candidates from TESS data 
assembled throughout the nominal and extended mission, covering more
than five years. The catalogues from \citet{Pedersen2019}, \citet{Antoci2019}, and \citet{Skarka2022}
are only the tip of the iceberg in discovery space for TESS
gravito-inertial asteroseismology.

\begin{acknowledgements}
  The research leading to these results has received funding from the
  KU\,Leuven Research Council (grant C16/18/005: PARADISE). CA
  acknowledges financial support from the Research Foundation Flanders
  (FWO) under grant K802922N (Sabbatical leave); she is grateful for
  the kind hospitality offered by CEA/Saclay during her sabbatical
  work visits in the spring of 2023. The authors are grateful to May
  Gade Pedersen and Joey Mombarg for providing their forward
  asteroseismic models in electronic format, to Alex Kemp for
  valuable comments on the manuscript prior to its submission, {
    and to the referee for constructive comments and encouragements to 
    expand the manuscript with more details.}
\end{acknowledgements}

\bibliographystyle{aa} 
\bibliography{GIW.bib} 

\end{document}